\documentclass[aps,reprint]{revtex4-1}
\usepackage{graphicx}
\usepackage{array}
\usepackage{amsmath}
\usepackage{lipsum}
\usepackage{bm}
\usepackage{xcolor}
\usepackage{natmove}
\usepackage{multirow}
\usepackage{soul}

\begin{document}
\title{Optical dependence of electrically-detected magnetic resonance in lightly-doped Si:P devices}

\author{Lihuang Zhu}
\author{Kipp J. van Schooten}
\author{Mallory. L. Guy}
\author{Chandrasekhar Ramanathan}
\email[]{chandrasekhar.ramanathan@dartmouth.edu}

\affiliation{Department of Physics \& Astronomy, Dartmouth College, Hanover, New Hampshire 03755, USA}

\date{\today}

\begin{abstract}

Electrically-detected magnetic resonance (EDMR) provides a highly sensitive method for reading out the state of donor spins in silicon.  The technique relies on a spin-dependent recombination (SDR) process involving dopant spins that are coupled to interfacial defect spins near the Si/SiO$_2$ interface.  To prevent ionization of the donors, the experiments are performed at cryogenic temperatures and the mobile charge carriers needed are generated via optical excitation.  The influence of this optical excitation on the SDR process and the resulting EDMR signal is still not well understood.   Here, we use EDMR to characterize changes to both phosphorus and defect spin readout as a function of optical excitation using: a 980 nm laser with energy just above the silicon band edge at cryogenic temperatures; a 405 nm laser to generate hot surface-carriers; and a broadband white light source.   EDMR signals are observed from the phosphorus donor and two distinct defect species in all the experiments.  With near-infrared excitation, we find that the EDMR signal primarily arises from donor-defect pairs, while at higher photon energies there are significant additional contributions from defect-defect pairs.  The optical penetration depth into silicon is also known to be strongly wavelength dependent at cryogenic temperatures.  The energy of the optical excitation is observed to strongly modulate the kinetics of the SDR process.  Careful tuning of the optical photon energy could therefore be used to control both the subset of spin pairs contributing to the EDMR signal as well as the dynamics of the SDR process.
\end{abstract}

\maketitle 

\section{Introduction}

Spin-based quantum phenomena at the nanoscale hold promise for the development of quantum-enhanced sensing and qubit-based computing architectures.  In order to fully realize this potential, however, it is necessary to interface these phenomena to macroscopic scales.  Isolated semiconductor dopants and defects offer long coherence times, robust and accurate quantum control and can be integrated into realistic device geometries. Some of the more intensively studied systems are nitrogen- and silicon-vacancy centers in diamond \cite{Staudacher2013,Rogers2014,Mamin2013}, various defects in silicon carbide \cite{Falk2015,Calusine2016}, as well as group V donors in silicon such as phosphorus \cite{Steger2012,Gumann2014,Hoehne2015}, arsenic \cite{Franke2015} and bismuth \cite{Wolfowicz2013}. 

The most widely studied dopant in silicon is phosphorus (Si:P) which has a single naturally occuring spin-1/2 isotope $^{31}$P.  The coherence times of this system are extremely long, up to seconds for electron spins and tens of minutes for the nuclear spins \cite{Tyryshkin2012,Saeedi2013}, among the longest reported for spins in solids. Furthermore, the use of silicon offers the advantage of mature fabrication methods and ease of integration with commercial nanoelectronics, making it a nearly ideal system in which to engineer scalable quantum technologies \cite{Kane1998,Hill2015}, albeit at cryogenic temperatures ($< 20$ K) to prevent ionization of the donor atoms.

Electrically-detected magnetic resonance (EDMR) of Si:P samples was first observed by Schmidt and Solomon over 50 years ago \cite{Schmidt1966} and has become an important tool for magnetic resonance of donors in micro- and nanoscale silicon devices due to its high sensitivity \cite{Stich1995,McCamey2006}. EDMR in Si:P has been used to electrically detect donor spin states \cite{Stegner2006}, and to readout an ensemble nuclear spin memory with extremely long lifetimes ($> 100$ s) \cite{McCamey2010}.  Silicon EDMR has been integrated with photoconductive AFM into a scanning probe microscope \cite{Klein2013} and has been used to detect the protons from water adsorbed onto the silicon surface \cite{Dreher2015}. 

\begin{figure}[ht]
  \includegraphics[width=0.5\textwidth]{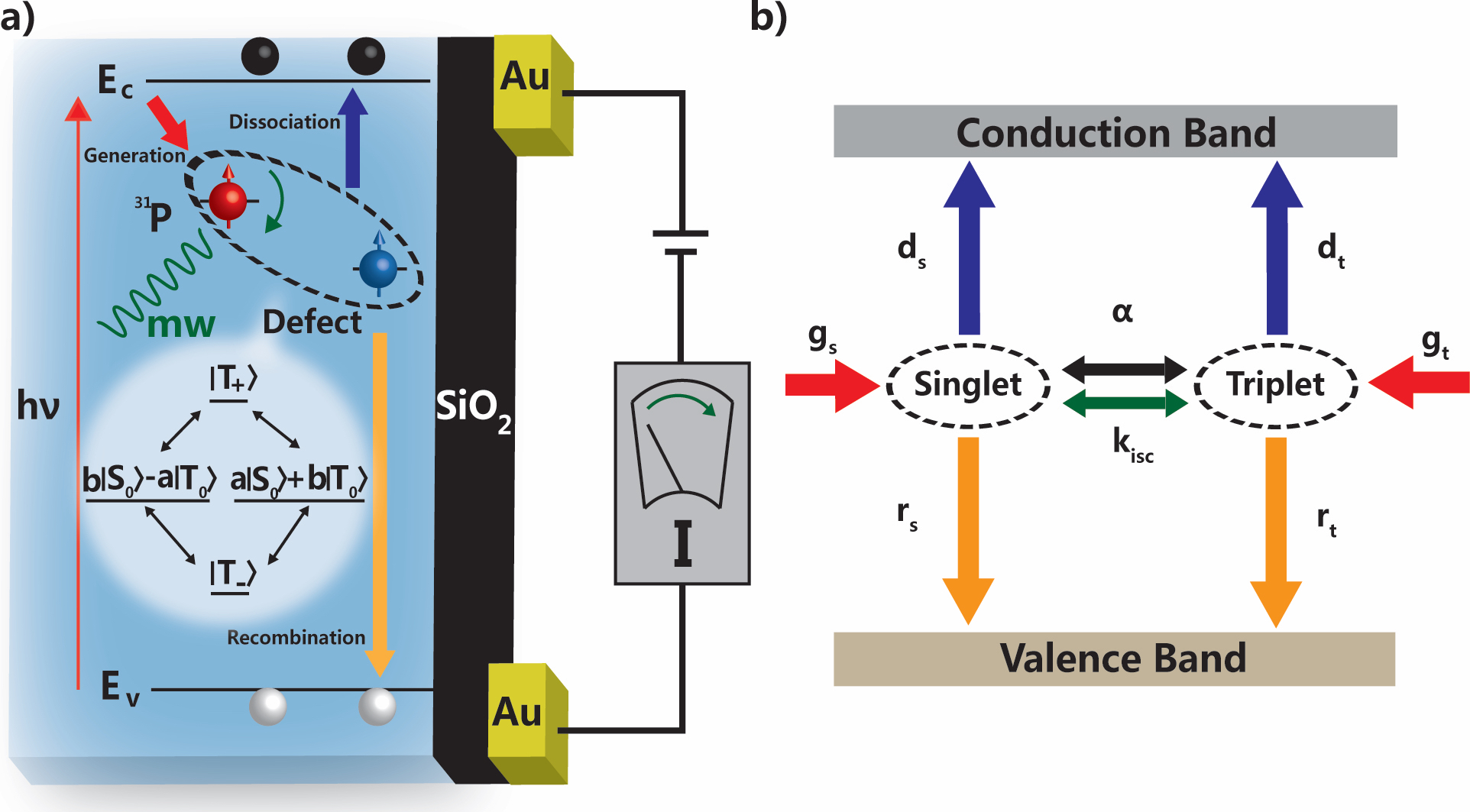}
  \caption{ (a) Schematic of spin-dependent recombination (SDR) at the Si/SiO$_2$ interface of a Si:P device. Optically excited electrons in the conduction band can get trapped at phosphorus donor sites that are coupled to adjacent interfacial defect spins.  The permutation symmetry of the coupled spin pair determines its recombination rate. Resonantly exciting one of the spins changes the symmetry and thus the recombination probability, resulting in a change in the electrical current through the device.  The eigenstates for the coupled spin pair are also shown. 
  (b)  A simple EDMR rate model proposed by Lee {\em et al}.\ \cite{Lee2012}.  The singlet and triplet pairs are created at rates $g_s$ and $g_t$ ($g_t = 3g_s$); dissociate at rates $d_{s}$ and $d_{t}$; and recombine at rates $r_{s}$ and $r_{t}$.  Microwave excitation induces a spin-mixing process at rate $\alpha$ while $k_{\mathrm{isc}}$ describes the inter-system crossing between the singlet and triplet manifolds.
 }
 \label{fig:fig1}
\end{figure}

Multiple mechanisms are known to mediate the spin-dependent transport that enables EDMR in different experimental configurations \cite{Kaplan1978,Honig1970,Ghosh1992}.  At low fields ($< 1$ T) where the longest coherence times have been observed, the dominant mechanism is spin-dependent recombination (SDR), where the recombination of a pair of spins depends on their spin permutation symmetry. Resonant excitation of either spin changes this symmetry, modulating the current through the device.  In Si:P, such spin pairs can be formed by phosphorus donors and paramagnetic defects located at the Si/SiO$_2$ interface, between pairs of defects or even between pairs of donors at higher doping concentrations \cite{Stegner2006,Hoehne2010,Hoehne2013}.  At cryogenic temperatures and low-doping concentrations, optical excitation is used to generate the free carriers necessary for EDMR. The influence of this optical excitation on the SDR rates and the observed EDMR signal is still not well understood.  While most EDMR experiments have used white light sources for the optical excitation \cite{Stegner2006,McCamey2006,McCamey2008,Morishita2009,Hoehne2010,Lee2010}, light emitting diodes \cite{Hoehne2013,Hoehne2015} and laser excitation \cite{Stich1995} have also been used.  At cryogenic temperatures the optical penetration depth of light into silicon is known to be strongly wavelength dependent \cite{Macfarlane1958}.  
Thus both the kinetic energy and the spatial distribution of the photo-excited carriers changes with wavelength. Broadband optical excitation, for example, generates both hot carriers along with near-band-edge carriers -- with differing spatial distributions.

Here, we investigate the wavelength dependence of the EDMR signal in a Si:P device, using three different optical sources: a 980 nm laser whose energy is just above the band edge of silicon at cryogenic temperatures, a 405 nm laser to generate hot surface-carriers, and a broadband tungsten-halogen lamp white light source.  
 With near-infrared excitation, we find that the EDMR signal primarily arises from donor-defect pairs, while at higher photon energies there are significant additional contributions from defect-defect pairs.  Using frequency modulated (FM) continuous-wave (CW) EDMR we measure the modulation frequency and microwave power dependence of the EDMR signal for each optical excitation and show that the optical excitation energy can strongly modulate the kinetics of the SDR process.  Careful tuning of the optical photon energy could therefore be used to control both the subset of spin pairs contributing to the EDMR signal as well as the dynamics of the SDR process.

\section{Spin Dependent Recombination}

\noindent If a sample of Si:P is irradiated with above gap light at low temperatures, a steady-state photocurrent is generated where the optical excitation rate is balanced by the carrier recombination rate.  If any of the recombination pathways is spin-dependent, a resonant excitation of the spins can modulate the recombination rate and transiently change the current through the sample, a mechanism proposed by Kaplan, Solomon and Mott \cite{Kaplan1978}.

Figure~\ref{fig:fig1}(a) illustrates the basic EDMR experiment in Si:P.  Shallow phosphorus donor electrons near the Si/SiO$_2$ interface interact with adjacent (deep) paramagnetic defects present at the interface via either dipolar or exchange interactions.  
The four energy eigenstates for the spin pair are $|T_{+}\rangle = |\uparrow\uparrow\rangle$, $|T_{-}\rangle = |\downarrow\downarrow\rangle$, and the two admixed states $|1\rangle = 
a|S_0\rangle + b|T_0\rangle$ and $|2\rangle = 
b|S_0\rangle - a|T_0\rangle$, where $|T_0\rangle$ and $|S_0\rangle$ are the $m_s=0$ triplet and singlet states.  For a strongly-coupled pair, the states $|1\rangle$ and $|2\rangle$ become the singlet state $|S_0\rangle$ and the triplet state $|T_0\rangle$ ($a=1,b=0$), while for very weak coupling they become the product states $|\uparrow\downarrow\rangle$ and $|\downarrow\uparrow\rangle$ ($a=b=1/\sqrt{2}$). 

Since silicon has low spin-orbit coupling, the recombination process is spin-preserving, resulting in faster recombination rates for states with singlet character compared to states with triplet character.  During steady state optical excitation,  most pairs are pumped into the states $|T_+\rangle$ or $|T_-\rangle$, since all the states are generated at the same rate (by non-geminate carriers) but $|1\rangle$ and $|2\rangle$ can recombine relatively quickly, given their singlet content.  Resonant microwave excitation of either spin can induce transitions from states $|T_+\rangle$ and $|T_-\rangle$ to states $|1\rangle$ or $|2\rangle$, resulting in a change in current.   

Lee {\em et al}.\ proposed a two-component (singlet/triplet) kinetic model to describe the signal dependence observed in CW EDMR experiments that takes into account the competing generation, recombination and dissociation processes \cite{Lee2012}.  Figure \ref{fig:fig1}(b) illustrates the key parameters of this model.  Under optical excitation, spin pairs are randomly generated in each of the four above configurations with equal probability,
so that the singlet and triplet generation rates $g_s$ and $g_t$ are related by $g_t = 3g_s$.  The singlet and triplet populations dissociate at rates $d_{s}$ and $d_{t}$, releasing an electron to the conduction band, and  recombine at rates $r_{s}$ and $r_{t}$, when one of the electrons in the pair recombines with a hole in the valence band.  Transitions between the singlet and triplet manifolds can be induced by either microwave excitation or via relaxation processes.  To lowest order, the microwave-induced transition rate $\alpha$ is proportional to the microwave power, while relaxation to thermal equilibrium populations is assumed to occur at a rate $k_{\mathrm{isc}}$ via inter-system crossover.   Assuming a simple on-off amplitude modulation scheme, they derived a set of coupled differential equations describing the changes to the free carrier populations and the current through the device.  The key equations describing the model are shown in Appendix~A.

While the SDR mechanism for phosphorus donors is believed to primarily be mediated by mid-gap dangling-bond P$_\mathrm{b0}$ defects \cite{Poindexter1981, Stesmans1998,Stegner2006,Hoehne2010}, previous EDMR measurements have measured E' defects \cite{Lee2010} as well as P$_\mathrm{b1}$ defects and a central donor pair resonance \cite{Hoehne2010}.  It was recently shown that EDMR in Si:P is primarily sensitive to those donors located within roughly the first 20 nm of the Si/SiO$_2$ surface \cite{Suckert2013}. The properties of a single donor-defect pair were also recently characterized using scanning probe techniques \cite{Ambal2016}.

\section{Experimental Setup}

\begin{figure}[t]
  \includegraphics[width=0.4\textwidth]{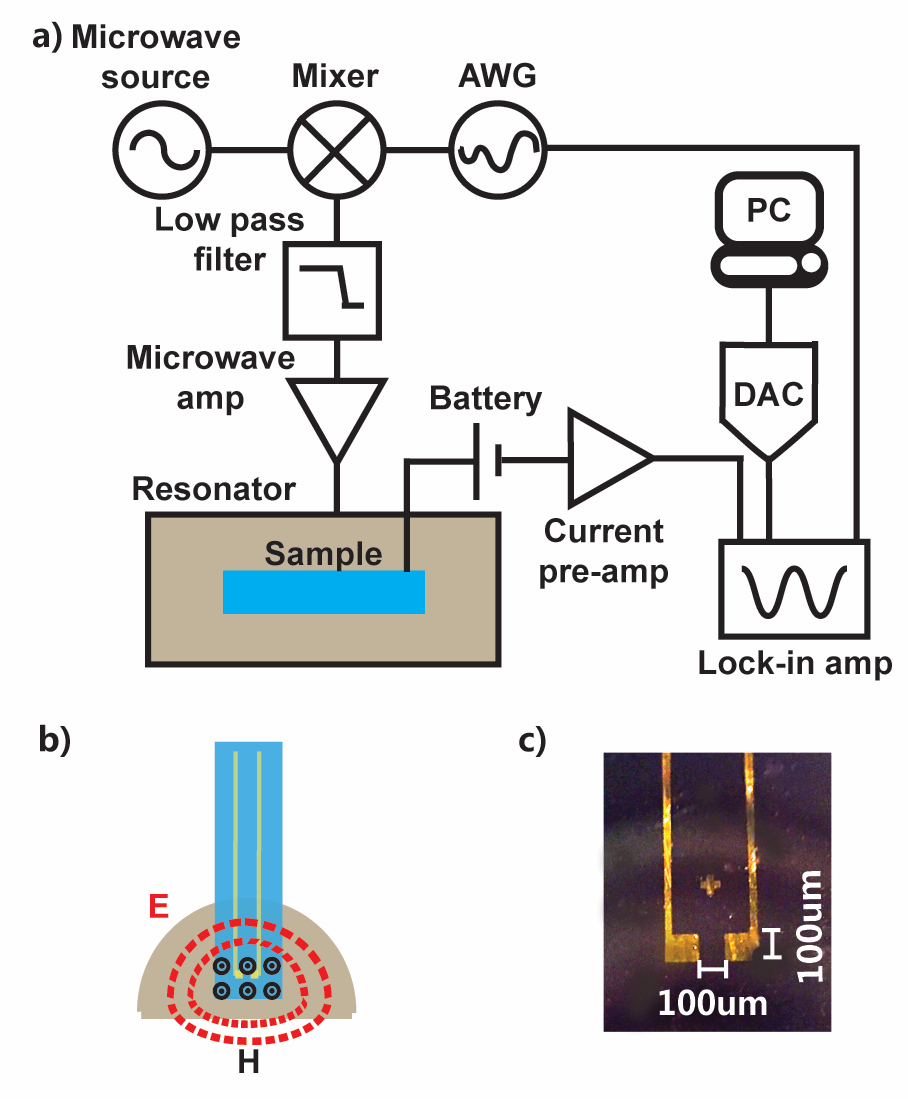}
  \caption{a) Block diagram of relevant portions to the experimental setup. b) Configuration on the sample (blue box) and the half-cylindrical resonator showing the electric ($\vec{E}$) and magnetic ($\vec{H}$) field orientations; c) Microscope image of the device.  The cross in the center is an alignment marker.
} 
 \label{fig:fig2} 
\end{figure}

\noindent Figure~\ref{fig:fig2}(a) shows a schematic of the experimental setup used.  The static magnetic field was generated by a 3-inch diameter electromagnet (Spectromagnetic Model 1019). A microwave synthesizer (QuickSyn FSW-0020) provided a constant carrier frequency of 2.596 GHz, which was mixed (Marki T3-06LQP) with a discrete, numerically-generated, frequency-modulation (FM) or amplitude-modulation (AM) waveform loaded into a high-frequency arbitrary waveform generator (Tektronix AWG7052). Low-pass filtering (Mini-Circuits VLF-2250+) was used to attenuate the upper sideband and carrier components by approximately 20 dB.  The microwaves were then amplified by 30 dB (Mini-Circuits amplifiers ZX60-V62 and ZX60-6019 in series), before being transmitted to the sample.  The microwaves were coupled to the sample with a lab-built, low quality-factor ($Q$), stripline-fed dielectric antenna mounted on the cold-finger of a continuous-flow Janis optical cryostat. 

Figure~\ref{fig:fig2}(b) shows the mode structure of the half-cylinder dielectric antenna used in the experiment (described in more detail in Appendix~B).  The relative alignment of the sample and antenna was set to minimize RF electric-field ($\vec{E}$) coupling to the electric current ($\vec{I}$) through the device  ($\vec{E} \perp \vec{I}$), since such a coupling  can excite microwave-induced currents that could mask the spin-dependent current changes.  The stripline-fed dielectric resonator had a 3 dB bandwidth of 7 MHz, centered at 2.596 GHz, resulting in a $Q$ of 371 at $T=4.2$ K.   

A battery and resistor network were used to provide a constant bias current, $I_{0}$, for a given optical illumination of the Si:P device.  The current was fed to an SRS 570 current amplifier, which also compensated for the constant current bias.  With 405 nm and white light excitation, the signals were measured in low-noise mode with a sensitivity setting of $10^{-6}$ A/V, while the high-bandwith mode and a $10^{-7}$ A/V sensitivity were used with the 980 nm excitation.  No additional filtering was performed in the current preamplifier. 
The output of the current amplifier was connected to an SRS 830 lock-in amplifier, to which the FM (or AM) waveform was input as a reference, and whose resulting output was digitized using a National Instruments NI-USB-6361 DAQ.  The time constant on the SR 830 was set to 100 ms in all the experiments described here.

The sample used in the experiment was fabricated on a commercial silicon on insulator (SOI) wafer (Ultrasil Corporation). The lightly phosphorus doped wafer had a device resistivity of 1-4 $\Omega\cdot$cm in the $<$100$>$ orientation, which corresponds to a phosphorus doping concentration of $1.2-5.0\times 10^{15}$ cm$^{-3}$.  This is significantly lower than the $10^{16}-10^{17}$ cm$^{-3}$ phosphorus donor concentrations used in most previously reported EDMR experiments \footnote{While Stich, {\em et al}.\ used samples with a donor concentration of $8\times 10^{14}$ cm$^{-3}$, they were unable to see an EDMR signal without irradiating the sample with 2 MeV electrons \cite{Stich1995}, thereby generating bulk donor-defect pairs}, where exchange interactions between the donors begins to become significant \cite{Cullis1975}.  The sample was mounted with the wafer parallel to the magnetic field.   

The 2.0$\pm$0.5 $\mu$m thick device layer is located on a 1 $\mu$m buried oxide layer. The 500$\pm$10 $\mu$m thick handle layer is boron doped with a resistivity of 10-20 $\Omega\cdot$cm in the $<$100$>$ orientation. The native oxide surface layer has a thickness  $<$10 nm. Gold contacts (100 nm) were thermally evaporated onto the surface, creating a $100\times$100 $\mu$m junction as shown in Figure~\ref{fig:fig2}(c) (Additional processing steps are described in Appendix~C).  This corresponds to an active device volume (assuming a sensitive depth of 20 nm) on the order of $2.0\times 10^{-10}$ cm$^3$ containing about $0.2-1.0\times 10^6$ donor electron spins. The typical surface density of both $P_{\mathrm{b0}}$ and E' defects is in the range of $10^{12}$ cm$^{-2}$ \cite{Lenahan1998,Herve1992,Takahashi1987}, leading to an estimate of about $10^8$ defect spins in the active device area. 

\begin{figure}[t]
  \includegraphics[width=0.4\textwidth]{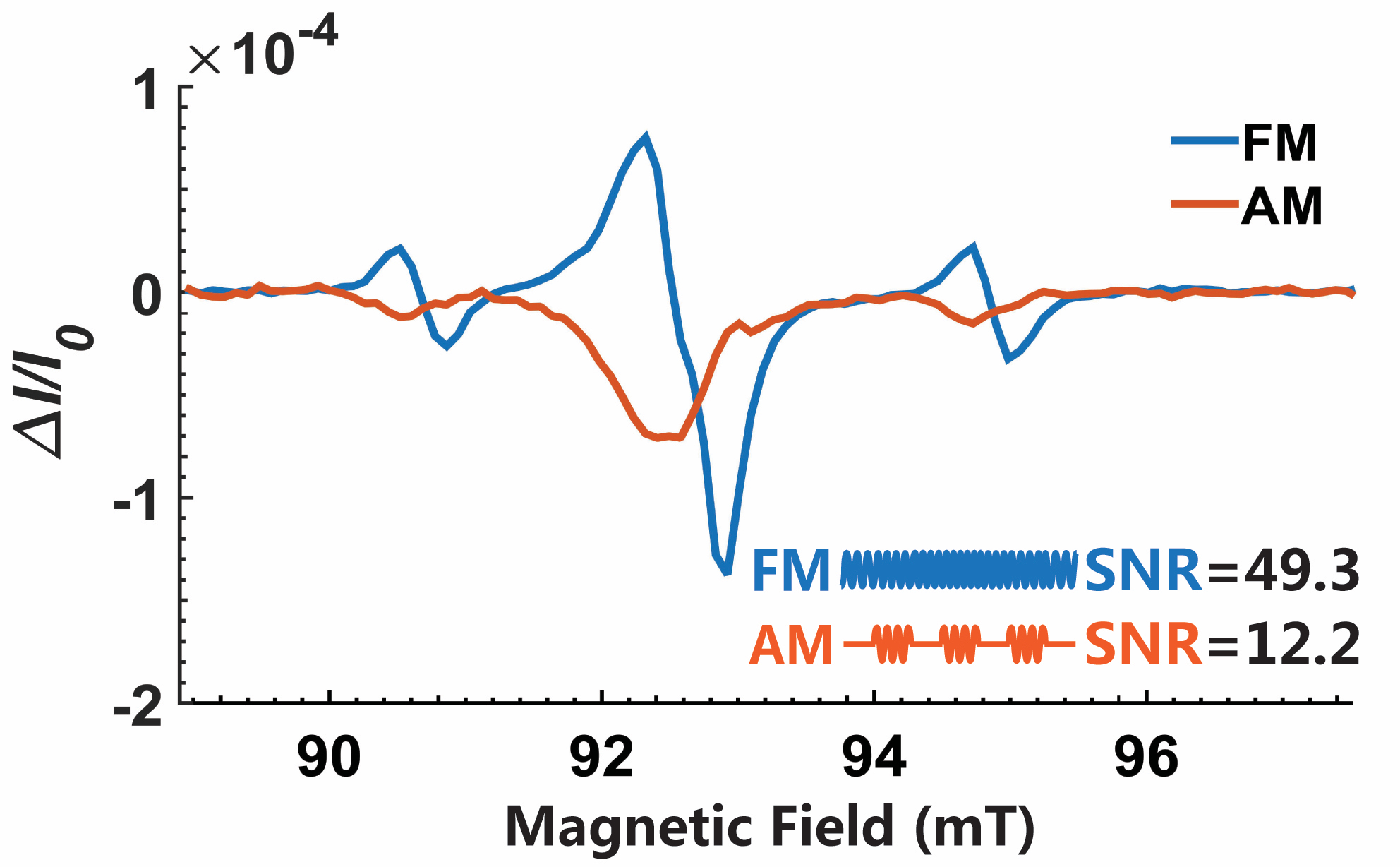}
  \caption{The CW EDMR spectrum obtained using FM and AM microwave modulation. Each spectrum is acquired at 4.2 K under tungsten halogen lamp broadband illumination and 1 kHz modulation frequency. The maximum power of the 2.596 GHz microwave was 3.16 W.  The signal-to-noise ratios (SNR) reported here are for a single scan.  An SNR improvement of $\sim 4$ is obtained for FM over AM. All other experimental parameters were kept the same in the two experiments.}
  \label{fig:fig3}
\end{figure}

\section{Results and Discussion}
\subsection{Microwave Modulation}

\noindent Although magnetic field modulation has traditionally been used for lock-in detection of CW-ESR and CW-EDMR, the use of small modulation coils (both to minimize inductance and due to space constraints) can lead to larger magnetic field inhomogeneities \cite{Crosser2010}. Additionally, vibrations due to Lorentz forces and direct inductive pickup of the field modulation by the electrical leads can lead to increased noise in EDMR signals.  AM microwave can also be used to detect EDMR, but the microwave-induced currents in the device electrode also pick up the modulation frequency and can mask the true EDMR signal as described in the previous section.  For FM microwave modulation, the $\vec{E}$-field coupling to the sample can be minimized since the microwave induced current is constant over the range of modulation frequencies, so that the modulation envelope is only transferred into the signal under magnetic resonance conditions. Here, a triangular envelope was used for the FM frequency variation, with the maximum frequency deviation set to 12 MHz, slightly larger than the measured resonator bandwidth.

Figure \ref{fig:fig3} shows a comparison between FM EDMR spectra (blue) and AM EDMR spectra (red) using a 1 kHz modulation frequency under white light excitation.  The central peak is due to surface defects while the two outer lines correspond to the 4.2 mT ($117.54$ MHz) hyperfine split lines of the phosphorus donors ($g=1.9985$) \cite{Stegner2006,Morishita2009}.  The transconductance gain of the current preamplifier was used to calculate the fractional current change from the measured signal voltage.  Note that, although FM EDMR results in derivative lineshape spectra, AM EDMR does not.  The peak microwave power delivered to the sample was kept constant at 3.16 W in both experiments.  Part of the difference in peak signal intensity between the two spectra is likely due to the lower average microwave power (a factor of 3 for a symmetric triangular waveform) in the AM experiment.  However, the signal-to-noise ratios (SNR) measured in the two experiments differ by a factor of 4, indicating a superior sensitivity for FM over AM EDMR.  Typical resonant changes in device current of $\Delta I/I_{0} = (10^{-4}-10^{-5})$ are observed.

\subsection{Optical Selection of Spin-Pair Species}

\begin{figure}[h]
  \includegraphics[width=0.35\textwidth]{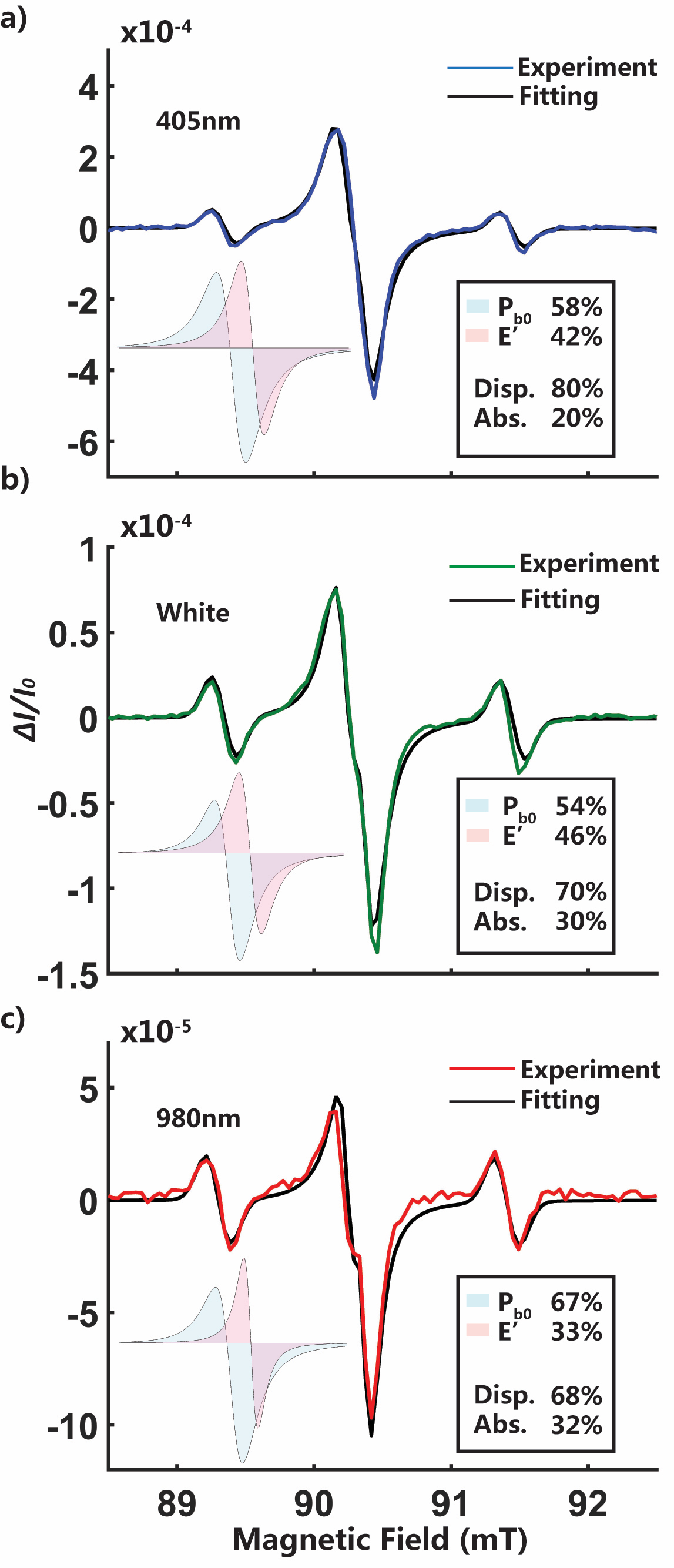}
  \caption{\label{fig:fig4} FM EDMR spectrum measured under optical excitation at a) 405 nm, b) white light and c) 980 nm. In each panel, the colored line shows the recorded spectrum while the black line shows a spectral fit. The 117.4 MHz hyperfine split phosphorus peaks were used to calibrate the field, with the $g$-factor of the phosphorus peak set to $g$=1.9985. The center defect peak was fit to the sum of two Lorentzian lines, one with a $g$-factor of 2.0058 (assigned to Pb$_0$) and the other with a $g$-factor of 2.0002 (assigned to E'). The peak with $g$-factor 2.0058 has a mix of dispersive and absorptive lineshapes. These spectra were collected at 4.2 K using a 1 kHz modulation frequency.   The microwave power used in these experiments was 3.16 W.}
\end{figure}

Figures~\ref{fig:fig4}(a), (b), and (c) show the EDMR spectra recorded using a 25 mW 405 nm laser ($3.06134$ eV; Edmond Optics 59562), a 6 W broadband white light source (OceanOptics LS-1-LL), and a 200 mW 980 nm laser ($1.26514$ eV; ThorLabs L980P200) respectively. Under the same bias conditions, the induced photocurrent ($I_0$) in the sample was 5 $\mu$A for the blue laser, 40 nA for the infra-red laser, and 1 $\mu$A for the tungsten halogen lamp. For bias voltages under 5 V, the leakage current in the dark was negligible. The microwave power used in these experiments was 3.16 W, which is sufficient to saturate the EDMR spectra, as shown later.

The (peak-to-peak) fractional current change for the phosphorus donors ($\Delta I_{\textrm{Phos}}/I_0$) changes from $9.7\pm1.8 \times10^{-5}$ at 405 nm illumination to about $3.8\pm0.3 \times 10^{-5}$ for 980 nm and $4.7\pm0.3 \times 10^{-5}$ for white light illumination.  The intensity of the central defect peak depends much more strongly on the optical excitation, with $\Delta I_{\textrm{Def}}/I_0$ changing from $7.6\pm0.2 \times 10^{-4}$ at 405 nm to $2.20\pm0.04 \times 10^{-4}$ under white light and $1.40\pm0.02 \times 10^{-4}$ at 980 nm.  The ratio between the two signals $\Delta I_{\textrm{Def}}/\Delta I_{\textrm{Phos}}$ changes from $7.8\pm1.0$ at 405 nm to $4.6\pm0.1$ with white light and $3.6\pm0.1$ at 980 nm.  Table \ref{tab:tab1} summarizes these results.  This change in the ratio between the two signals suggests that additional defect-defect interactions are contributing to the EDMR signal under 405 nm excitation.  The area of the defect peak is greater than the sum of the two hyperfine split phosphorus peaks in all the experiments. 

At the low donor concentrations used here, we do not expect donor pair resonances to arise. However multiple donor-defect and defect-defect EDMR signals are likely to be present. The figures also show the result of a spectral fit.  The 117.4 MHz hyperfine-split phosphorus peaks were used to calibrate the field, with the $g$-factor of the phosphorus peak set to $g$=1.9985. The center defect peak was fit to the sum of two Lorentzian lines.  One of the peaks has a $g$-factor of 2.0002 which is close to the reported value ($g$ = 2.0005) of deep hole oxide trap E'  defects \cite{Lenahan1998}. The other peak has a $g$-factor of 2.0058 which is intermediate between the $g$-values reported for P$_\mathrm{b0}$ ($g_1 = 2.0015$ - parallel to (111); $g_2 = 2.0080$; $g_3 = 2.0087$ - parallel to (011)) and P$_\mathrm{b1}$  ($g_1 = 2.0012$; $g_2 = 2.0076$ - parallel to (111); $g_3 = 2.0052$ - parallel to (011)) at the orientation used in the experiment \cite{Poindexter1981,Lenahan1998}.   We have labeled this the P$_\mathrm{b0}$ defect since this is the most-commonly observed defect peak in EDMR.  The shifts in the observed $g$-factor are most likely due to errors in sample alignment with the field. The peak is also observed to have a mixed absorptive and dispersive character.  Zevin and Suss have shown that such distortions of the line-shape can be caused by the microwaves passing through conducting metallic or semiconducting layers \cite{Zevin1986}.  The dispersive component could arise from defect spins in the buried oxide layer.  The  distortion in the line-shape is more obvious under 980 nm excitation where the optical penetration is the greatest. The contribution of the E' signal also drops while that of the P$_\mathrm{b0}$ signal increases for the long wavelength excitation.  This suggests that the observed E' defects are primarily located on the top surface while the P$_\mathrm{b0}$ defects are present at both the surface and buried oxide layers.

The width of the phosphorus peaks ($\sim3.5$ G) and the P$_\mathrm{b0}$ peak ($\sim2.7$ G) remained relatively unchanged in the different experiments.  The width of the E' peak changed from $\sim2.9$ G for 405 nm and white light excitation to $\sim1.9$ G for 980 nm excitation.  This is consistent with a weaker perturbation of the surface E' spins with long wavelength excitation. 

\begin{center}
\begin{table*}[t]
   \renewcommand{\arraystretch}{1.5}
    \begin{tabular}{ || l | >{\centering}p{1cm}| >{\centering}p{1cm}| >{\centering}p{1cm} |>{\centering}p{1cm}| >{\centering}p{1cm}| >{\centering}p{1cm}| >{\centering}p{1.8cm}| >{\centering}p{1.8cm}| >{\centering}p{1.9cm}|| }
    \hline
 \multirow{2}{*}{Source} & P$_0$ &P$_\mathrm{I}$ & P$_\mathrm{2 \mu m}$ & I$_0$ & $\lambda$ & P$_\mathrm{20 nm}$ & $\Delta I_{\textrm{Phos}}/I_0$ & $\Delta I_{\textrm{Def}}/I_0$  & {$\Delta I_{\textrm{Def}}/\Delta I_{\textrm{Phos}}$} \arraybackslash \\
 & (mW) & (mW) & (mW) & ($\mu$A) & ($\mu$m) & (mW) & ($\times 10^{-5})$ & ($\times 10^{-5})$ &  \\ \hline
  980 nm & 200 & 63.7 & 1.2 & 0.04 & 100 & 0.012 & $3.8\pm0.3$ & $14.0\pm0.2$ & $3.7\pm0.1$ \arraybackslash\\ \hline
   white & 6000 & 1910 & - & 1  & - & - & $4.7\pm0.3$ & $22.0\pm0.4$ & $4.7\pm0.1$ \arraybackslash\\ \hline
   405 nm & 25 & 8 & 8 & 5 & 0.12 & 1.2 & $9.7\pm1.8$ & $76.0\pm2.0$ & $7.8\pm1.0$ \arraybackslash \\ \hline      \end{tabular}
\caption{Optical dependence in the FM EDMR experiment. P$_0$ is the nominal optical power of the source; P$_\mathrm{I}$ is the optical power incident on the $100\times100$ $\mu$m device area (assuming a circular spot size with a 100 $\mu$m radius); P$_\mathrm{2\mu m}$ is the optical power deposited in the 2 $\mu$m device layer; I$_0$ is the steady-state light-induced photocurrent; $\lambda$ is the characteristic penetration depth for the optical excitation (inverse of the absorption coefficient); P$_\mathrm{20 nm}$ is the optical power deposited in the top 20 nm, where the SDR process dominates; $\Delta I_{\textrm{Phos}}/I_0$ is the fractional current change of the phosphorus donors; $\Delta I_{\textrm{Def}}/I_0$ is the fractional current change of the defect spins; $\Delta I_{\textrm{Def}}/\Delta I_{\textrm{Phos}}$ is the ratio of the current change for defects to the current change for the phosphorus.}
\label{tab:tab1}
\end{table*}
\end{center}
 
Given the nominal incident powers and taking literature values for silicon absorption coefficient at these wavelengths \cite{Green1995}, the calculated absorbed optical power over the device active volume ranges from 12 $\mu$W at 980 nm to 1.2 mW at 405 nm.  However, the induced steady-state photocurrent ($I_0$) is likely to pass uniformly through the entire 2 $\mu$m device layer for the 980 nm excitation, given the 100 $\mu$m penetration depth, but be more inhomogeneously distributed for the 405 nm excitation.  While the optical penetration is restricted to about 120 nm at this wavelength, the carriers are likely to diffuse through the entire 2 $\mu$m device layer.  However the surface contribution to the overall current will be significantly higher for the 405 nm excitation than for the 980 nm excitation.  This suggests that the fractional current changes could be made much larger using excitation in the infra-red if the current paths could be constrained to the surface, as has been done with the use of epitaxially-grown silicon layers \cite{Suckert2013}.

\begin{figure*}[t]
  \includegraphics[width=0.8\textwidth]{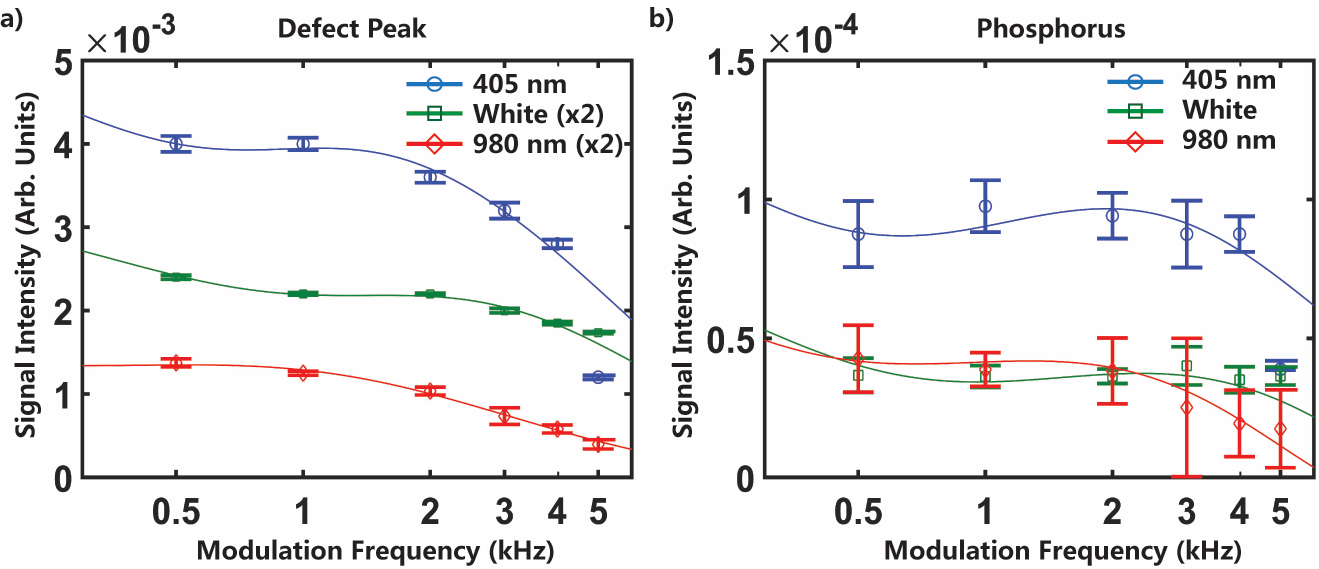}
  \caption{\label{fig:fig5} Modulation frequency dependence of the EDMR signal for the three different optical excitations for a) the defect signal, and b)the phosphorus donor signal. The microwave power was held constant at 3.16 W in these experiments.  The solid lines are simulations of the signal dependence predicted by the two-spin kinetic model shown in Figure~\ref{fig:fig1}(b).}
\end{figure*}

The excess energy of the incident photons relative to the silicon band-gap is rapidly dissipated through electron and phonon scattering that can significantly modify the kinetics of the SDR process.  At 4.2 K, silicon possesses two thresholds for indirect band-gap transitions, with the higher at 1.2135 eV \cite{Macfarlane1958}.  As a consequence, the types of carriers excited under each illumination varies widely. Excitation at 980 nm, just above the second phonon-mediated absorption threshold, generates relatively low-energy carriers, while 405 nm excitation leads to absorption enhancement of nearly three orders in magnitude \cite{Jellison1982}, generating hot carriers and increasing the phonon bath. The broadband white light source spans both regimes, while also exciting sub-band transitions such as donor-bound excitonic transitions, as have recently been exploited to perform bias-free EDMR experiments in isotopically-enriched silicon-28 samples \cite{Steger2012,Saeedi2013}. 

\begin{figure*}[t]
  \includegraphics[width=0.8\textwidth]{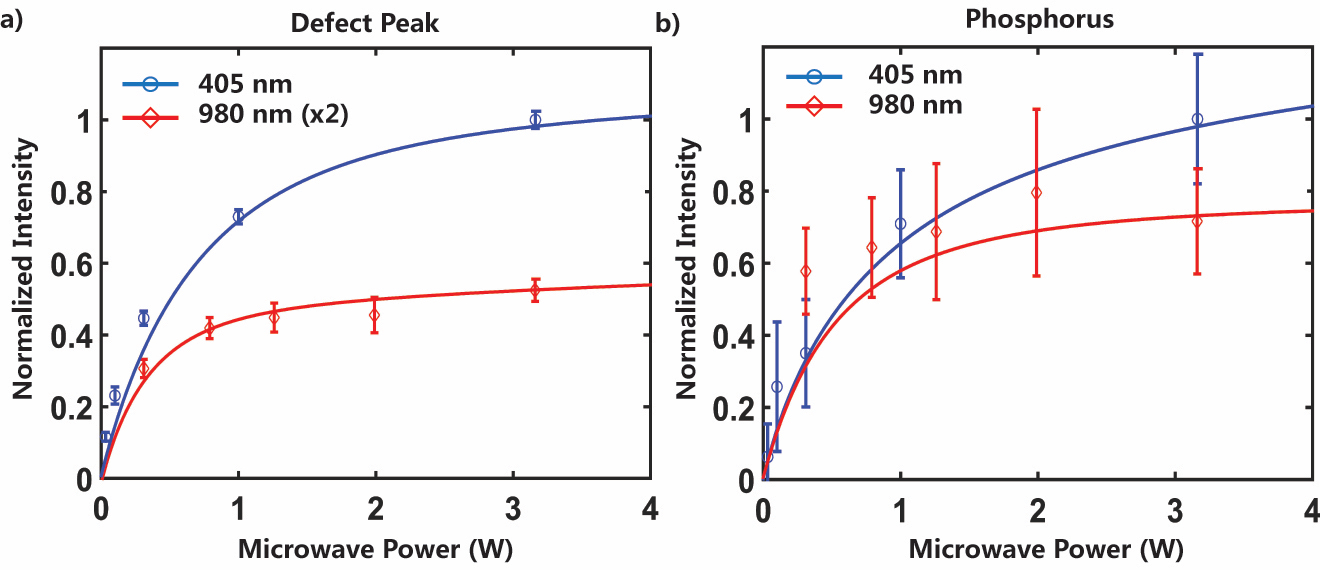}
  \caption{\label{fig:fig6}  Microwave power dependence of the EDMR signal for excitation with the 405 nm and 980 nm laser sources for a) the defect peak, and b) the phosphorus donor peak. The modulation frequency used was 1 kHz.  The solid lines are simulations of the signal dependence predicted by the two-spin kinetic model shown in Figure~\ref{fig:fig1}(b).}
\end{figure*}

\begin{center}
\begin{table*}[t]
    \renewcommand{\arraystretch}{1.5}
    \begin{tabular}{ || c | >{\centering}p{2cm} >{\centering}p{2cm} >{\centering}p{2cm}| >{\centering}p{2cm} >{\centering}p{2cm} >{\centering}p{2cm} || }
    \hline
  &\multicolumn{3}{c}{Defects} & \multicolumn{3}{|c||}{Phosphorus} \\ \hline
Source & 405nm & White & 980nm & 405nm & White & 980nm \arraybackslash \\ \hline
$k_{isc}$&$3.1\times10^{4}$&$2\times10^{4}$&$1\times10^{4}$&$1.4\times10^{4}$&$9.5\times10^{3}$&$8.7\times10^{3}$ \arraybackslash \\
$r_{s}$&$14.9\times10^{4}$&$7.6\times10^{4}$&$6.6\times10^{4}$&$4.6\times10^{4}$&$2.9\times10^{4}$&$2.8\times10^{4}$ \arraybackslash \\
$r_{t}$&$8.1\times10^{3}$&$7.8\times10^{3}$&$8.3\times10^{3}$&$5.5\times10^{3}$&$1.9\times10^{3}$&$1.6\times10^{3}$ \arraybackslash \\
$d_{s}$&$7.5\times10^{3}$&$5\times10^{3}$&$1\times10^{3}$&$4.9\times10^{3}$&$2.2\times10^{3}$&$1.9\times10^{3}$  \arraybackslash \\
$d_{t}$&$4.9\times10^{4}$&$4.5\times10^{4}$&$2\times10^{4}$&$2.9\times10^{4}$&$2.5\times10^{4}$&$2.4\times10^{4}$ \arraybackslash \\ \hline
    \end{tabular}
\caption{Fitting parameters used in Figure~\ref{fig:fig5} for different optical excitations. $\alpha=7.2\times10^{5}$ was used in all the experiments conducted with same microwave power. We set $g_{t}$=3$g_{s}$ in all the experiments, and used $g_{s} = 10^{24}$ for 405 nm excitation, $10^{23}$ for white light excitation and $10^{22}$ for 980 nm excitation. All parameters have units of s$^{-1}$.}
\label{tab:tab2}
\end{table*}
\end{center}

\subsection{Wavelength-Dependent Rate Changes}

\noindent In order to better connect to the changing kinetics of the SDR process, we measured the modulation frequency and microwave power dependence of the EDMR signal for each optical excitation.  Figure~\ref{fig:fig5}(a) and (b) show the modulation frequency dependence of the phosphorous donor and overall central defect signal intensities.  The current change is observed to decrease at higher modulation frequencies in all cases.  The change in EDMR with modulation frequency is an indirect probe of the SDR kinetics \cite{Hoehne2013,Lee2012}.  The solid lines in Figure~\ref{fig:fig5} show the simulated signal dependence predicted by the kinetic model of the EDMR process described earlier \cite{Lee2012}.  Note that while this model was developed for a simple on-off amplitude modulation of the EDMR signal, we are using it here to approximately describe the triangular frequency modulation signal measured in our experiments.  Table~\ref{tab:tab2} shows the parameters in these simulations.  Dreher {\em et al}.\  have reported singlet and triplet recombination time constants to be 15 $\mu$s and 2 ms respectively in Si:P \cite{Dreher2015}.  However the other rates for this system have not been measured to date.  Our initial estimate for these kinetic parameters were taken from Ref.\ \cite{Lee2012}.   We assumed these rates would not change by more than an order of magnitude, thus keeping the  general shape of the modulation dependence the same.  Appendix~D outlines the detailed data processing steps and the calculation of the error bars shown. 

In general, we see that almost all the electronic rates for both defects and phosphorus signals are higher for the 405 nm excitation experiment.  For the defect signal, the singlet recombination rate at 405 nm is a factor of 2 higher than the rate at 980 nm or with white light excitation. Overall the electronic recombination and dissociation rates for the defect signal are observed to be higher than for the phosphorus signal. However, the model fails to capture the signal decrease at the highest modulation frequency under 405 nm excitation.  This is probably due to the fact the observed signal arises from a number of different spin pairs, while the simulations are performed on a single pair.  The central defect signal could have contributions from P$_\mathrm{b0}$-P$_\mathrm{b0}$, P$_\mathrm{b0}$-E', E'-E', P$_\mathrm{b0}$-Phosphorus and E'-Phosphorus pairs.  Appendix~E shows the change in the different defect components as a function of modulation frequency.  It should be noted that these signals still represent the average behavior of multiple spin species, and could be partially correlated with each other.

Figures~\ref{fig:fig6}(a) and (b) show the microwave power dependence of the two components for the blue and red laser excitation, showing that the fractional current change initially increases with microwave power before saturating, as has been observed previously \cite{Stich1995}. To match the curves in Figure~\ref{fig:fig6} the parameter $\alpha$ was varied (assumed directly proportional to power) while all other parameters were kept fixed.

Care should be taken in interpreting the above changes in rate constant quantitatively, as Lee {\em et al}.\  have shown that a wide range of combinations of electronic rates can give rise to the same modulation frequency dependence \cite{Lee2012}.

\section{SUMMARY and OUTLOOK}

In summary, we have demonstrated high-sensitivity FM EDMR in lightly-doped Si:P devices, making comparative measurements on the optical excitation dependence of the EDMR spectra. We find that photon energies just above silicon's phonon-mediated absorption threshold lead to a spin-spin population dominated by dopant-defect pairs, while the generation of hot carriers greatly increases the population fraction of defect-defect pairs.   Two types of defect species were observed, which we ascribe here to P$_\mathrm{b0}$ and E' defects.  The contribution of an absorptive component to the EDMR signal from the P$_\mathrm{b0}$ defects suggests that a part of this signal arises from defects adacent to the buried oxide layer of the silicon-on-insulator sample.
The underlying cause of the observed wavelength-dependent changes can be at least partially understood in the context of dramatically different optical absorption cross-sections between the two excitation energy extremes. Optical absorption at the surface Si/SiO$_2$ interface is enhanced as the photon energy is increased, while the relative contribution of the buried oxide layer is more important at longer wavelengths. Additionally, the SDR rate kinetics are observed to change with the excitation source, possibly due to the amount of excess energy the photo-excited carrier dissipates during the capture process.

The tuning of surface spin-selectivity via optical excitation could enable the use of such silicon-based devices as quantum-enhanced surface-selective biochemical sensors. Demonstrations of this type of technology have been previously accomplished using NV centers in diamond for local nuclear magnetic resonance (NMR) detection of protons within nm$^3$ voxels \cite{Staudacher2013,Mamin2013}.  However, the difficulty in controlling the orientation of the NV axis in implanted centers makes it challenging to build NV-based sensor arrays with ordered site spacings below the optical diffraction limit. On the other hand, the ability to lithographically pattern structures on silicon surfaces could enable the design of sensor arrays which are highly scalable.  As Dreher {\em et al}.\ have shown previously, EDMR can be used to detect protons adsorbed onto the silicon surface, analogous to NMR measured by way of NV centers \cite{Dreher2015}.  This coupling between interfacial P$_\mathrm{b0}$ defects and surface nuclear spin species has also been observed in dynamic nuclear polarization experiments \cite{Cassidy2013,Guy2017}. In principle, it should be possible to resonantly detect any spin system -- electronic or nuclear -- that is coupled to the interfacial defect spins. Paramagnetic electronic states contributing directly to the SDR mechanism would be particularly attractive since their presence or absence could be immediately discerned through acquisition of a simple CW EDMR spectrum. In this case, optimizing optical excitation for surface-localized electronic generation would restrict EDMR readout to interface spin-states, enhancing SDR sensitivity to the current fraction arising from this region.

\begin{acknowledgments}

We thank Professor Christoph Boehme at the University of Utah for several helpful discussions during the initial setting up of the experiment.  We thank Dwayne Adams and Chris Grant for their help with designing and machining various components of the experimental setup.  This work was funded in part by the National Science Foundation under CHE-1410504. 

\end{acknowledgments}

\bibliography{newPRApplied}

\appendix

\begin{figure}[h]
\includegraphics[width=0.35\textwidth]{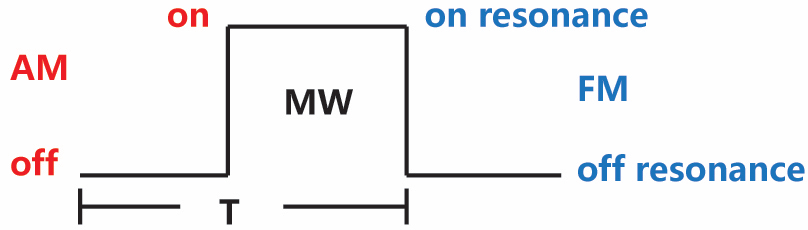}
\caption{\label{fig:supp1} The rate model was developed for on-off amplitude modulation of the microwaves. We have assumed that this is equivalent to an on-resonance / off-resonance frequency modulation of the microwaves, which should qualitatively mimic the behavior of the triangular frequency modulation used in our experiments.}
\end{figure}

\section{Spin Pair Rate Model}
\noindent A spin pair model described by Lee {\em et al.}\  was used to simulate the modulation frequency and power dependence shown in Figures~\ref{fig:fig5} and \ref{fig:fig6} \cite{Lee2012}.  
We can use two coupled rate equations to describe the changes to the number of singlet ($n_s$) and triplet ($n_t$) spins in the model of Figure \ref{fig:fig1}(b).  

\begin{eqnarray}
\frac{dn_{s}}{dt} &=&g_{s}-(d_{s}+r_{s})n_{s}+\alpha(n_{t}-n_{s})-k_{\mathrm{isc}}(n_{s}-Fn_{s})+ \nonumber \\
& & \hspace*{0.1in} k_\mathrm{isc}\left[n_{t}-(1-F)n_{t}\right]  \label{eq:app1}\\
\frac{dn_{t}}{dt} &=&g_{t}-(d_{t}+r_{t})n_{t}+\alpha(n_{s}-n_{t})-k_\mathrm{isc}(n_{t}-Fn_{t})+ \nonumber \\
& & \hspace*{0.1in} k_\mathrm{isc}\left[n_{s}-(1-F)n_{s}\right]
\label{eq:app2}\end{eqnarray}
where $g_{s},r_s,d_s,g_{t},r_t,d_t$ are the generation, recombination and dissociation rates for singlet and triplet spin pairs. $\alpha$ represents the microwave-induced transition rate between $n_{s}$ and $n_{t}$ while $k_\mathrm{isc}$ describes the inter-system crossing which restores the populations of $n_{s}$ and $n_{t}$ to thermal equilibrium.  $F$ is the Fermi-Dirac distribution function, $F = \left(1+e^{\frac{\Delta E}{kT}}\right)^{-1}$, which is set to 0.25 in the modeling results shown.

These two equations are solved for square-wave AM microwave modulation as shown in Figure~\ref{fig:supp1}, with $\alpha$ $\neq$ 0 when the microwaves are on and  $\alpha$ = 0 when the microwaves are off, resulting in 

\begin{eqnarray}
n_{s}^{\textrm{on}}(t) &=& A_{11}e^{-m_{11}t}+A_{21}e^{-m_{21}t}+n_{s}^{\textrm{on}}(ss) \hspace*{0.85in}\label{eq:app3} \\
n_{t}^{\textrm{on}}(t) &=& B_{11}e^{-m_{11}t}+B_{21}e^{-m_{21}t}+n_{t}^{\textrm{on}}(ss) \label{eq:app4}  \\
n_{s}^{\textrm{off}}(t) &=& A_{12}e^{-m_{12}(t-\frac{T}{2})}+A_{22}e^{-m_{22}(t-\frac{T}{2})}+n_{s}^{\textrm{off}}(ss) \label{eq:app5} \\
n_{t}^{\textrm{off}}(t) &=& B_{12}e^{-m_{12}(t-\frac{T}{2})}+B_{22}e^{-m_{22}(t-\frac{T}{2})}+n_{t}^{\textrm{off}}(ss) \label{eq:app6}
\end{eqnarray}

\noindent where $n_{s}^{\textrm{on}}$ and $n_{t}^{\textrm{on}}$ are the singlet and triplet populations when the MW pulse is on and $n_{s}^{\textrm{off}}$ and $n_{t}^{\textrm{off}}$ are the singlet and triplet population when the MW pulse is off. $n_{s/t}^{\textrm{on/off}}(ss)$ are the steady-state solutions of $n_{s/t}^{\textrm{on/off}}$, assuming that the modulation rate ($1/T$) is very low. 
The amplitudes $A_{ij}$, $B_{ij}$ and the time constants ($m_{ij}$) of the exponential functions depend on the electronic rates $\alpha, g_{s}, g_{t}, r_{s}, r_{t}, d_{s}, d_{t}$ and $k_{isc}$.  In order to solve for these, 8 boundary conditions are applied to Equations~\ref{eq:app3}--\ref{eq:app6}. The first four conditions represent the periodicity of the solution, namely, $n_{s}^{\textrm{on}}(0)$ = $n_{s}^{\textrm{off}}(T)$, $n_{t}^{\textrm{on}}(0)$ = $n_{t}^{\textrm{off}}(T)$, $n_{s}^{\textrm{on}}(T/2)$ = $n_{s}^{\textrm{off}}(T/2)$ and $n_{t}^{\textrm{on}}(T/2)$ = $n_{t}^{\textrm{off}}(T/2)$. The other four boundary condition are simply the fact that the only allowed population change in $n_{s}$ and $n_{t}$ are caused by generation, recombination, dissociation and the two spin mixing process.

The electrical signal is proportional to $d_{s}n_{s}$+$d_{t}n_{t}$, which leads to the in-phase and out-of-phase electrical signals from the lock-in amplifier  \cite{Lee2010}

\begin{widetext}
\begin{eqnarray}
I_{\mathrm{in}} &= & \frac{2m_{11}}{T}\left(r_{s}A_{11}+r_{t}B_{11}\right)\left[\frac{1-e^{-m_{11}T/2}\cos(l\pi)}{m_{11}^{2}+4l^{2}\pi^{2}/T^{2}}\right]+\frac{2m_{21}}{T}\left(r_{s}A_{21}+r_{t}B_{21}\right)\left[\frac{1-e^{-m_{21}T/2}\cos(l\pi)}{m_{21}^{2}+4l^{2}\pi^{2}/T^{2}}\right] + \\ 
& & \frac{2m_{12}}{T}\left(r_{s}A_{12}+r_{t}B_{12}\right)\left[\frac{\cos(l\pi)-e^{-m_{12}T/2}}{m_{12}^{2}+4l^{2}\pi^{2}/T^{2}}\right]+\frac{2m_{22}}{T}\left(r_{s}A_{22}+r_{t}B_{22}\right)\left[\frac{\cos(l\pi)-e^{-m_{22}T/2}}{m_{22}^{2}+4l^{2}\pi^{2}/T^{2}}\right] \nonumber 
\end{eqnarray}

\begin{eqnarray}
I_{\mathrm{out}} &= &\frac{4l\pi}{T^{2}}\left(r_{s}A_{11}+r_{t}B_{11}\right)\left[\frac{1-e^{-m_{11}T/2}\cos(l\pi)}{m_{11}^{2}+4l^{2}\pi^{2}/T^{2}}\right]+\frac{4l\pi}{T^{2}}\left(r_{s}A_{21}+r_{t}B_{21}\right)\left[\frac{1-e^{-m_{21}T/2}\cos(l\pi)}{m_{21}^{2}+4l^{2}\pi^{2}/T^{2}}\right] +  \\ 
& & \frac{4l\pi}{T^{2}}\left(r_{s}A_{12}+r_{t}B_{12}\right)\left[\frac{\cos(l\pi)-e^{-m_{12}T/2}}{m_{12}^{2}+4l^{2}\pi^{2}/T^{2}}\right]+\frac{4l\pi}{T^{2}}\left(r_{s}A_{22}+r_{t}B_{22}\right)\left[\frac{\cos(l\pi)-e^{-m_{22}T/2}}{m_{22}^{2}+4l^{2}\pi^{2}/T^{2}}\right] + \nonumber  \\
& &\left(r_{s}\Delta n_{s}(ss)+r_{t}\Delta n_{t}(ss)\right)\left[\frac{\cos(l\pi)-1}{l\pi}\right] \nonumber 
\end{eqnarray}
where $\Delta n_{s}(ss)$ = $n_{s}^{\textrm{off}}(ss)$-$n_{s}^{\textrm{on}}(ss)$ and $\Delta n_{t}(ss)$ = $n_{t}^{\textrm{off}}(ss)$-$n_{t}^{\textrm{on}}(ss)$.  In our experiment, we use the magnitude output of the lock-in amplifier, instead of measuring the in-phase and out-of-phase signal changes, so the measured signal intensity is proportional to 
\begin{equation}
S =\sqrt{I_{\mathrm{in}}^{2} + I_{\mathrm{out}}^{2}}.
\end{equation}
\end{widetext}

\begin{figure}[b]
\includegraphics[width=0.45\textwidth]{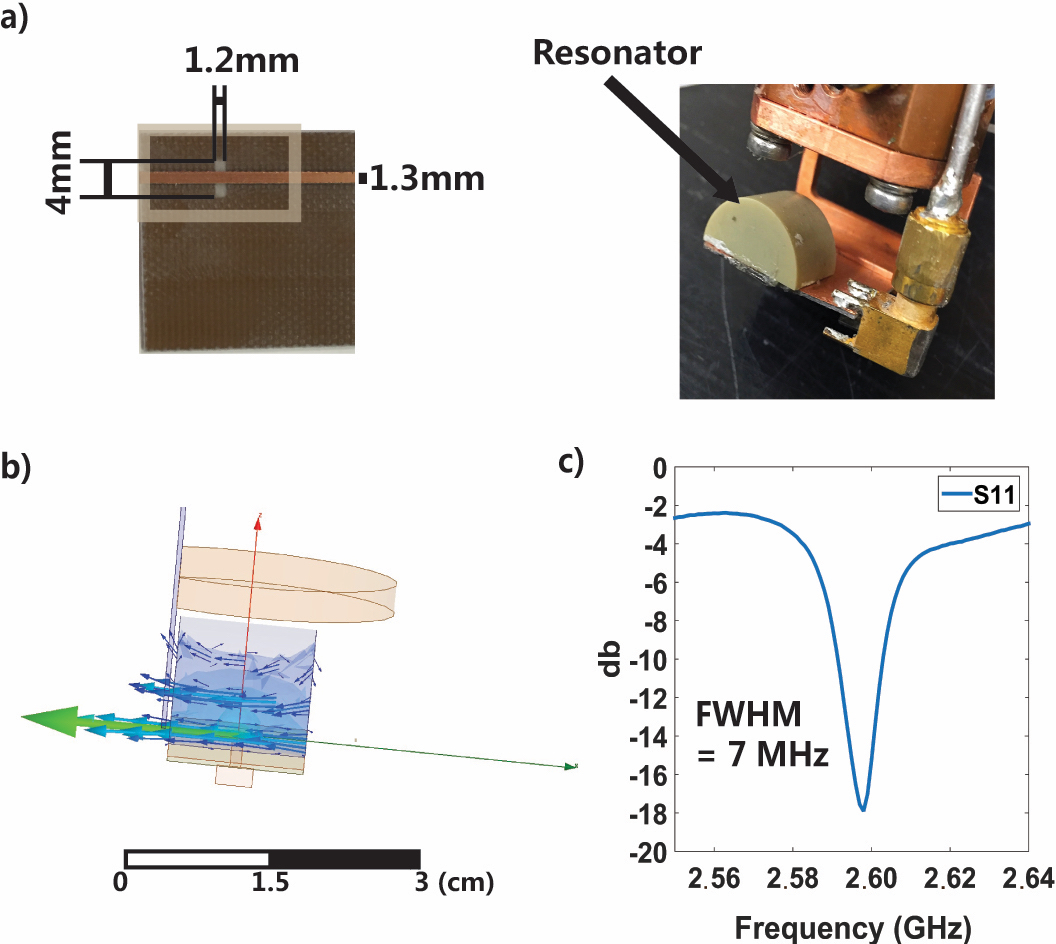}
\caption{\label{fig:supp2} a) (left) Microstrip-line and slot used to couple microwaves to the dielectric resonator.  (right) Photo of the dielectric resonator. b) High frequency electromagnetic field simulation software (ANSYS HFSS) simulation for the dielectric resonator at 2.596 GHz. The arrows indicate magnetic field vector. The figure on the right shows a schematic of the mode structure. c) The S$_{11}$ parameter of the resonator measured at 4.2 K with a network analyzer.}
\end{figure}

\newpage

\section{Dielectric Resonator}

\noindent The half-cylindrical dielectric resonator was purchased from TCI ceramics. This dielectric constant of this resonator is 81.0 $\pm$2. The dimensions of this half-cylindrical resonator are shown in Figure~\ref{fig:supp2}(a). Microwaves are coupled to the dielectric resonator through a strip-line fabricated on a two-sided printed circuit board (PCB).  A small slot is cut just above the strip line on the opposite side of the PCB and the dielectric resonator is centered over the slot. The TE$_{01\delta}$ mode is excited at 2.596 GHz at 4.2 K.  Figure~\ref{fig:supp2}(b) shows an electromagnetic field simulation (ANSYS HFSS) of the dielectric resonator at 2.596 GHz, and a schematic of the mode structure.  Figure~\ref{fig:supp2}(c) shows the measured S$_{11}$ parameter of the resonator at 4.2 K, corresponding to a $Q$-factor of 370.9.

\begin{figure}[t]
\includegraphics[width=0.5\textwidth]{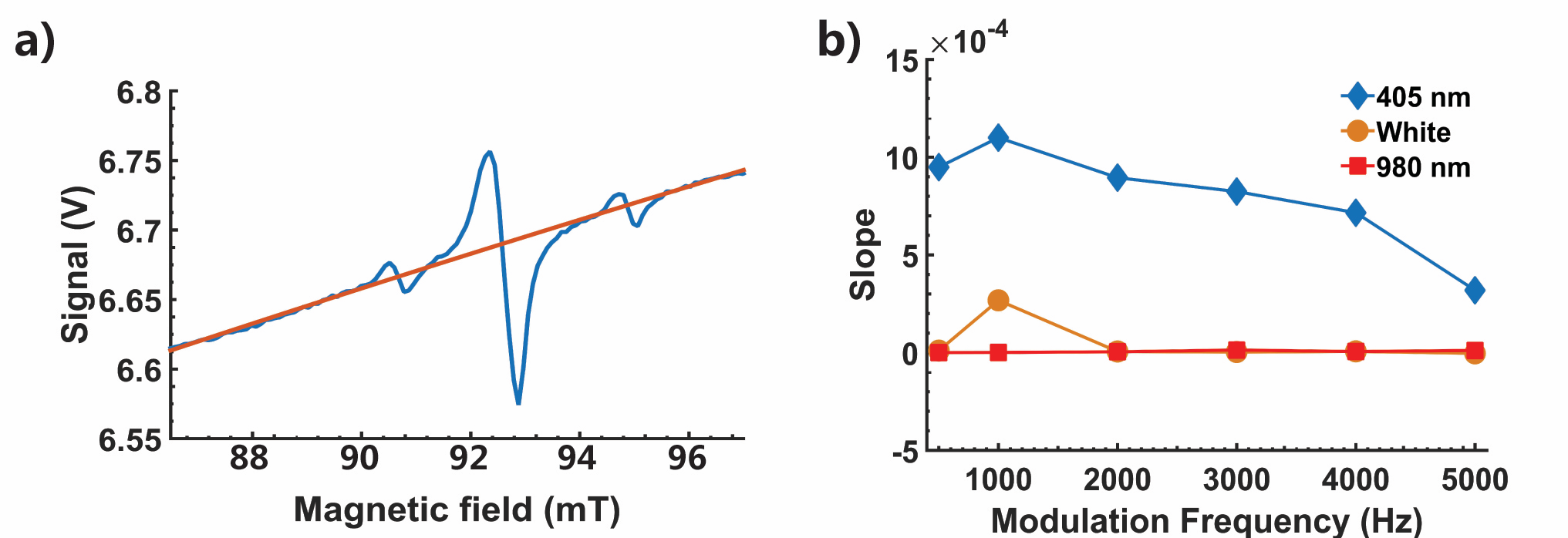}
\caption{\label{fig:supp3} a) Raw CW EDMR spectrum (blue) and first order polynomial fit used for baseline correction (red). b) Slope of the first order polynomial equation obtained as a function of modulation frequency for each of the three light sources.}
\end{figure}

\section{Device Fabrication}
\noindent The wafer was first immersed in 6:1 buffered oxide etch (BOE) solution for 5 min to remove the native oxide layer on top of the silicon device layer. A 1.5 $\mu$m thick layer of S1813 Shipley photoresist was then spin-coated onto the sample as soon as possible, followed by a 3 minute soft bake at 100 degrees Celsius. The features for the metal contacts were defined by exposure to 26 mW/cm$^2$ 405 nm light for 15 seconds using a mask aligner. The sample was then developed in Microposit MF319 developer for 1 minute, which was followed by a 5 minute hard bake at 100 Celsius.  
\begin{figure}[h]
\includegraphics[width=0.32\textwidth]{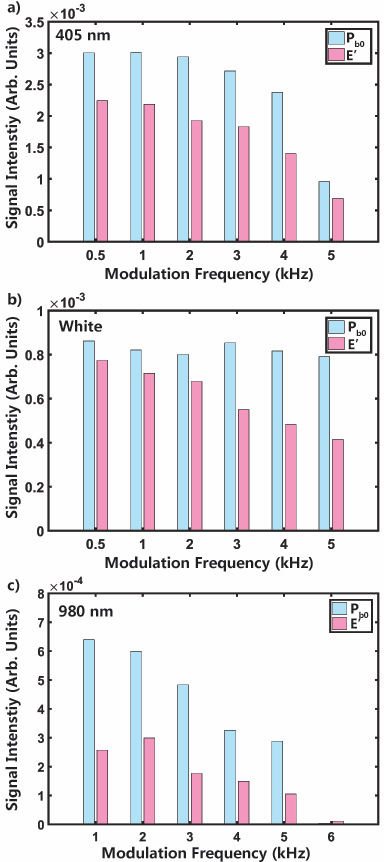}
\caption{Change in intensity of the two defect components under different modulation frequency and light excitation, obtained from the spectral fits.}
\label{fig:supp4}
\end{figure}

\newpage

\section{Data Analysis}
\noindent Figure~\ref{fig:supp3}(a) shows the raw EDMR data, illustrating the presence of a linear baseline.  In order to correct for this, we fit the baseline of the measured spectra with a first order polynomial equation and subtracted this from the data, resulting in the flat baselines seen in Figures~\ref{fig:fig2} and \ref{fig:fig3}.  Figure~\ref{fig:supp3}(a) also shows the fit used for the baseline correction.

Figure~\ref{fig:supp3}(b) shows the slopes of the linear fits obtained as a function of modulation frequency for each of the three optical excitation schemes, showing that the baseline correction did not significantly interfere with our analysis.   For the 405 nm laser signal, the slope of the linear fit shows a similar trend when compared to the signal intensity.  However, no such dependence is observed for the white light source and the 980 nm excitation. We currently do not know the origin of the baseline signal. However one possible explanation is that this slope is related to the magneto-resistance discovered in lightly doped phosphorus silicon \cite{Porter2012}.  

The signal intensity in the main manuscript is calculated from the resonance peak area after base line correction. The experimental spectra are scaled by lock-in amplifier and current preamplifier settings. The error bars shown in Figures~\ref{fig:fig5} and \ref{fig:fig6} were calculated using the standard deviation of the baseline (following subtraction of the linear fit).

\section{Spectral Fits for Modulation and Power Dependence}
\begin{figure}[h]
\includegraphics[width=0.32\textwidth]{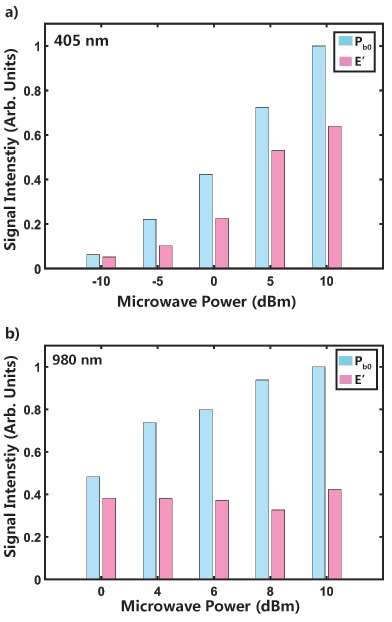}
\caption{Change in intensity of the two defect components with applied microwave power, obtained from the spectral fits.}
\label{fig:supp5}
\end{figure}

\noindent We performed two-component fits for the defect spectra measured under different modulation frequency and microwave power excitations.  The modulation frequency dependence is shown in Figure~\ref{fig:supp4}, while the microwave power dependence is shown in Figure~\ref{fig:supp5}.
The modulation frequency dependence of both components is similar to that of the total signal for the monochromatic excitations at 405 nm and 980 nm. With white light excitation, it appears that the main modulation dependence arises from E' defects.  The microwave power dependence of the two components follows the overall signal at 405 nm, but at 980 nm it appears that the $E'$ defect signal is independent of microwave power.  As noted earlier, care should be taken in interpreting these results as some of the defect signal also arises from E'-P$_\mathrm{b0}$ pairs, which results in correlated signals.

\end{document}